\documentclass[sn-mathphys]{sn-jnl}

\jyear{2021}%

\theoremstyle{thmstyleone}%
\newtheorem{theorem}{Theorem}
\newtheorem{proposition}[theorem]{Proposition}%

\theoremstyle{thmstyletwo}%

\theoremstyle{thmstylethree}%

\def\cK{\mathcal{K}}
\newcommand{\be}{\begin{equation}}
\newcommand{\ee}{\end{equation}}

\newcommand{\bea}{\begin{eqnarray}}
\newcommand{\eea}{\end{eqnarray}}

\newcommand{\beq}{\begin{equation}}
\newcommand{\eeq}{\end{equation}}

\def\cZ{\mathcal{Z}}
\def\cN{\mathcal{N}}

\newcommand{\mC}{{\mathbb C}}

\newcommand{\IR}{{\mathbb R}}
\newcommand{\IZ}{{\mathbb Z}}

 \def\cC{{\cal C}}

\newcommand{\cM}{\mathcal{M} }

\newcommand{\g}{ \gamma}
\newcommand{\s}{ \sigma }

\newcommand{\Dim}{\text{Dim}}

 \newcommand{\Rib}{ {\rm Rib}}
\newcommand{\id}{{\rm id}}
\newcommand{\cO}{\mathcal{O} }

\newcommand{\bdel}{ {\boldsymbol{\delta}} }

\newcommand{\pairs}{ { \rm pairs} } 
\newcommand{\singlets}{ {\rm singlets} } 

\begin{document}

\title[Kronecker coefficients from algebras of bi-partite ribbon graphs]{Kronecker coefficients from algebras of bi-partite ribbon graphs }

\author[1,3]{\fnm{Joseph } \sur{Ben Geloun}}
\email{bengeloun@lipn.univ-paris13.fr}

\author*[2,4]{\fnm{Sanjaye} \sur{Ramgoolam}}\email{s.ramgoolam@qmul.ac.uk}

\affil[1]{\orgdiv{Laboratoire d'Informatique de Paris Nord UMR CNRS 7030}, \orgname{Universit\'e Paris 13}, \orgaddress{\street{ 99 avenue J.-B. Clement}, \city{Paris}, \postcode{93430}, \state{Villetaneuse}, \country{France}}}

\affil[2]{\orgdiv{School of Physical and Chemical Sciences}, \orgname{Queen Mary University of London}, \orgaddress{\street{Mile End Road}, \city{London}, \postcode{E14 6HT}, 
\country{United Kingdom}}}

\affil[3]{\orgdiv{International Chair in Mathematical Physics
and Applications}, \orgname{ICMPA--UNESCO Chair}, 
 \postcode{072 B.P. 50}, \state{Cotonou}, \country{Benin}}

\affil[4]{\orgdiv{School of Physics and Mandelstam Institute for Theoretical Physics}, \orgname{ University of Witwatersrand}, \orgaddress{\street{Wits2050}, \city{Johannesburg}, 
 \country{South Africa}}}


\abstract{Bi-partite ribbon graphs arise in organising the large $N$ expansion of correlators in random  matrix models and in the enumeration of observables in random tensor models. There is an algebra $\mathcal{K}(n)$, with basis given by bi-partite ribbon graphs with $n$ edges, which is useful in the applications to matrix and tensor models. The algebra $ \mathcal{K}(n)$ is closely related to symmetric group algebras and has a matrix-block decomposition related to Clebsch-Gordan multiplicities, also known as Kronecker coefficients, for symmetric group representations. Quantum mechanical models which use $\mathcal{K}(n)$ as Hilbert spaces can be used to give combinatorial algorithms for computing the Kronecker coefficients.  
}



\maketitle

\section{Introduction}\label{sec1}

Kronecker coefficients are widely studied in mathematics from many points of view (e.g. symmetric polynomials, complexity theory, combinatorics). In this contribution to the volume on non-commutativity and physics, motivated by work on tensor models in physics, we review how the structure of a  class of non-commutative, associative algebras $\cK(n)$ (one for each integer $n$)  leads to new algorithms for computing Kronecker coefficients. These algorithms have a geometrical interpretation in terms of sub-lattices of a lattice whose basis vectors are labelled by bi-partite ribbon  graphs. Bi-partite ribbon  graphs themselves are related to holography in a fundamental way (which relates quantum field theory on one geometry to string theory on another geometry). The story we present has, as central figures,  the algebra $\cK(n)$ and  stringy  geometrical structures (lattices, holography). This can be viewed as part of a new direction in the study of non-commutative geometric structures in mathematical physics, which encompasses stringy holography as well as more direct approaches to the emergence of geometry from algebras in traditional  non-commutative geometry \cite{Connes,Madore,Majid}. The standard mathematical approach to  Kronecker coefficients is to think of them as the structure constants of the commutative fusion ring of representations of  symmetric groups. Bringing the non-commutative associative algebra  $\cK(n)$  to bear on the Kronecker coefficients  evidences the idea pioneered in traditional non-commutative geometry,  that underlying commutative mathematical structures, there are deeper non-commutative physical constructions which carry interesting hidden information.

Following the classic work of 't Hooft \cite{tHooft} it has been understood that the combinatorics of large gauge theories with gauge symmetries such as $U(N), SO(N), Sp(N)$ simplifies in the large $N$ limit. This simplification is related to the emergence of string world-sheet combinatorics in the $1/N$ expansion of physical observables, and underlies examples of gauge-string duality such as the AdS/CFT correspondence \cite{malda}. These simplifications are based on the appearance of double-line diagrams or equivalently ribbon graphs in large $N$ computations. Ribbon graphs are also related to the mathematics of holomorphic maps between two-dimensional surfaces, and in particular a special class of such maps called Belyi maps. This leads to simple mathematical models of gauge-string duality based on the correspondence between correlators of invariant matrix polynomials in matrix models (the gauge theory) and the counting of Belyi maps (which can be viewed as a combinatorial topological string theory) \cite{dMRBelyi}.  

The solution of counting problems for tensor model observables  can also be formulated in terms of ribbon graph counting \cite{BenGeloun:2020yau,JoSan1}. Specifically,
we consider a complex tensor variable $ \Phi_{ijk}$, with $ i , j, k \in \{ 1, 2, \cdots , N \} $ which transforms in the three-fold tensor product  $( V_N \otimes V_N \otimes V_N) $ 
of the fundamental representation $V_N$ of $U(N)$. The space of polynomials of degree $n$ in $\Phi$ and degree $n$ in the complex conjugate $\bar \Phi$ contains a subspace of $U(N)$ invariants of dimension, which somewhat non-trivially, turns out to be equal to the number of  bi-partite ribbon graphs with $n$ edges, for $ N \ge n$. It was also understood \cite{JoSan2}  that there is an associative algebra $\cK(n)$ with basis labelled by these bi-partite ribbon graphs. The dimension of the algebra, equivalently the number of bi-partite ribbon graphs with $n$ edges, is equal to the sum of squares of Kronecker coefficients : 
\bea 
\Dim ( \cK(n) ) = \sum_{ R_1 , R_2 , R_3 \vdash n } ( C ( R_1 , R_2 , R_3 ) )^2  
\eea
for all triples $R_1, R_2 , R_3 $ of Young diagrams with $n$ boxes. The algebra $\cK(n)$ has a Fourier basis $Q^{ R_1, R_2 , R_3 }_{ \tau_1 , \tau_2 } $  labelled by triples of Young diagrams along with a pair of multiplicity indices $\tau_1 , \tau_2$ which each range over
 $\{ 1, 2, \cdots , n \}$ \cite{PCA2016,JoSan2}.   The Fourier basis is defined using Clebsch-Gordan coefficients of the symmetric group $S_n$   and gives the Weddernurn-Artin decomposition of $\cK(n)$ as a direct sum of matrix algebras, where the matrix algebras exist for triples of Young diagrams having non-vanishing Kronecker coefficient. The papers \cite{DR1706,DGT1707,IMM1710} have related results on tensor model observables with a similar algebraic perspective. 
 
 In the paper \cite{BenGeloun:2020yau}, we observed that the matrix-subspace of $\cK(n)$ associated with a triple of Young diagrams can be obtained as the eigenspace of a specified set of central elements  in $\cK(n)$, acting on $\cK(n)$ by multiplication. This led to the realization of the square of the Kronecker coefficient for any triple $(R_1, R_2 , R_3)  $ as the number of linearly independent null vectors of a certain combinatorially constructed integer matrix. Integer matrix algorithms give a constructive combinatorical algorithm for arriving at a basis in the space of null vectors.  The construction has a natural interpretation in terms of quantum mechanics in a Hilbert space spanned by the bi-partite ribbon graphs with $n$ edges.   Furthermore, considering an involution operator on $\cK(n)$,  we arrive at the construction of the Kronecker coefficient itself, and therefore propose
 an answer to the old question of Murnaghan on a combinatorial interpretation of Kronecker coefficient \cite{MurnaghanOnReps,StanleyConjecture}. The complexity of the combinatorial algorithm is a left as a interesting problem for the future.

 In section \ref{sec:MatBel} we describe the combinatorial model of gauge-string duality arising from complex matrix model correlators. In section \ref{sect:CandAlg} we explain the quantum mechanics on the algebra of bi-partite ribbon graphs. In section \ref{HNF-algo} we explain the construction of the integer matrix for every triple, whose null space has a dimension given by the square of the Kronecker coefficient for that triple.  Further considerations lead us to the construction of several other pertinent  integer sub-lattices in the lattice of bi-partite ribbon graphs. An interesting corollary is the identity (\ref{sumsinglets}) giving the sum of Kronecker coefficients for triples of Young diagrams with $n$ boxes, in terms of the number of ribbon graphs invariant under an involution $S$ defined in section \ref{HNF-algo}. Finally we note that the use of combinatorial structures from string theory  and quantum gravity, notably combinatorial topological string theory, to address  mathematical group theory questions has also more recently found applications in the proof of integrality properties of  partial character sums for general finite groups \cite{IDFCTS,CTSGTA}, providing an alternative to the Galois theoretic proof  for these sums.

\section{ Bi-partite graphs and matrix model correlators  }\label{sec:MatBel}

Consider a Gaussian complex matrix model  of a complex matrix of size $N$ with  partition function 
\be
\cZ = \int [dZ] e^{ - { 1 \over 2 } tr Z Z^{ \dagger} }
\ee
$[dZ]= \prod_{ i , j } dZ^i_j d \bar Z^i_j $ is the standard measure on $N^2$ complex variables.  
Using standard formulae for multi-variable Gaussian integration, 
we find that the  quadratic  expectation value  is 
\begin{align} 
\langle Z^{i}_{ j}  ( Z^{ \dagger} )^k_l \rangle = { 1 \over \cZ} 
\int [dZ] e^{ - { 1 \over 2 } tr Z Z^{ \dagger} } Z^i_j (Z^{\dagger})^k_l = \delta^i_l \delta_j^k
\end{align} 
The complex matrix model measure is invariant under transformations by  matrices $U$ which are in the unitary group $U(N)$. The complex matrix model finds application in describing the half-BPS sector of the $\cN =4$ super-Yang Mills theory \cite{BBNS}\cite{CJR}. Holomorphic gauge invariant polynomial functions of $Z$  of degree $n$ correspond to half-BPS quantum states of scaling dimension $n$. These can be parametrised using permutations in $S_n$  
\bea 
 \cO_{ \sigma } ( Z ) = \sum_{ i_1 , \cdots , i_n }  Z^{ i_1}_{ i_{ \sigma(1) } } Z^{ i_2}_{ i_{ \sigma(2) } } \cdots  Z^{ i_n }_{ i_{\sigma (n) } }  
 = \prod_{ a =1}^{n} ( {\rm{tr}} Z^a)^{ C_a ( \sigma ) } 
\eea
where $C_{ a } ( \sigma )$ is the number of cycles of length $a$ in $\sigma$. It is easy to verify that $\cO_{ \sigma } (  Z ) = \cO_{ \gamma \sigma \gamma^{-1} } ( Z ) $ for any $ \gamma \in S_n$. The correlator of a holomorphic and an anti-holomorphic operator is defined by the integral 
\bea 
\langle \cO_{ \sigma_1} ( Z ) \cO_{ \sigma_2 } ( Z^{ \dagger} ) \rangle 
 = { 1 \over \cZ} 
\int [dZ] e^{ - { 1 \over 2 } tr Z Z^{ \dagger} } \cO_{ \sigma_1 } ( Z ) 
\cO_{ \sigma_2} ( Z^{ \dagger} )  \, . 
\eea
This is calculated to be \cite{CJR,BrownCMMD}
\bea 
 && \langle \cO_{ \sigma_1} ( Z ) \cO_{ \sigma_2 } ( Z^{ \dagger} ) \rangle  
 = \sum_{ \sigma_3 \in S_n } \sum_{ \gamma \in S_n } \delta ( \sigma_1 \gamma \sigma_2 \gamma^{-1} \sigma_3  ) N^{ C_{ \sigma_3 } }    \cr 
&& = { n! \over \vert \cC_1 \vert \vert \cC_2 \vert   }  
\sum_{ \sigma_1' \in \cC_1 } \sum_{ \sigma_2' \in \cC_2 } \sum_{ \sigma_3 \in S_n }  \delta ( \sigma_1' \sigma_2' \sigma_3 ) N^{ C_{ \sigma_3 } } \, . 
\eea 
Here $ \cC_1 $ is the conjugacy class of $\sigma_1 $, $\cC_2$ is the conjugacy class of $\sigma_2$ and $C_{ \sigma_3 }$ is the number of cycles in the permutation $ \sigma_3$. We have used $ \vert\cC_1 \vert , \vert \cC_2 \vert $ to denote the sizes of the conjugacy classes $\cC_1 , \cC_2$. We can also separate out the sum over $\sigma_3$ into all the conjugacy classes, labelled by $\cC_3$ in $S_n$ : 
\bea 
 \langle \cO_{ \sigma_1} ( Z ) \cO_{ \sigma_2 } ( Z^{ \dagger} ) \rangle  
= { n! \over \vert \cC_1 \vert \vert \cC_2 \vert   }  
\sum_{ \sigma_1' \in \cC_1 } \sum_{ \sigma_2' \in \cC_2 } \sum_{ \cC_3} \sum_{ \sigma_3 \in \cC_3  }  \delta ( \sigma_1' \sigma_2' \sigma_3 ) N^{ C (\cC_3) } \, . 
\eea
In the last expression we have used $C(\cC_3)$ for the number of cycles in any permutation belonging to the conjugacy class $\cC_3$. 

It is convenient to normalise the observables as follows : 
\bea\label{corrWS}  
&& N^{-n}  { N^{  C ( \cC_1 ) + C ( \cC_2 ) }  \vert \cC_1 \vert \vert \cC_2 \vert \over n ! n!  } 
  \langle \cO_{ \sigma_1} ( Z ) \cO_{ \sigma_2 } ( Z^{ \dagger} ) \rangle  \cr 
  && =\sum_{ \cC_3 } \left ( { 1 \over n! } \sum_{ \sigma_1' \in \cC_1 } 
  \sum_{ \sigma_2' \in \cC_2 } \sum_{ \sigma_3 \in \cC_3 } 
\cC  \delta ( \sigma_1' \sigma_2' \sigma_3 ) N^{ C ( \cC_1 ) + C ( \cC_2 ) + C ( \cC_3 ) - n  } \right ) 
\eea
The sum for fixed conjugacy classes $\cC_1 , \cC_2, \cC_3 $ has a nice geometrical interpretation in terms of holomorphic maps from a two-dimensional surface to a sphere, with three branch points on the sphere. In the inverse image of the branch points, the branching structure is described by the conjugacy classes $\cC_1 , \cC_2 , \cC_3$. The genus $h$ of the surface is given by 
\bea 
(2-2h )  = C ( \cC_1 ) + C ( \cC_2 ) + C ( \cC_3 ) - n 
\eea
The formula \eqref{corrWS} shows that the naturally normalised matrix model correlators can be interpreted as observables in a topological string with sphere (complex projective plane)  target, where the string path integral is  localised on holomorhic maps with three branch points, which can be chosen to be $0,1, \infty $. Such maps are distinguished maps of interest in number theory, called Belyi maps. There are also nice combinatorial objects, namely bi-partite ribbon graphs associated with these Belyi maps. These can be thought of as the inverse image of the interval $[0,1]$. For a review of the link between bi-partite ribbon graphs, and references to the original literature, we refer the reader to \cite{dMRBelyi}. A good textbook discussion of bi-partite ribbon graphs and Belyi maps is in \cite{LZBelyi}.

\section{Counting, algebra and quantum mechanics of tensor model observables}
\label{sect:CandAlg}

\noindent{\bf Counting of rank $d$ tensor model observables.}
The counting  of rank $d$ complex tensor observables or  tensor invariants,  containing $n$ copies of a complex tensor variable $\Phi$  and its conjugate $\bar \Phi$,   has been performed in \cite{JoSan1}. 
The enumeration method of tensor invariants used therein is based on the counting of equivalence classes
in Cartesian products of the symmetric group $S_n$ of $n$ elements, generated by certain subgroup actions.   We describe it here at rank $d=3$, using the ``gauge-fixed formulation'' from \cite{JoSan1,JoSan2}, while  the generalisation to any $d$ is straightforward. 

\begin{figure}[h]\begin{center}
     \begin{minipage}[t]{.8\textwidth}\centering
\includegraphics[angle=0, width=6cm, height=1.8cm]{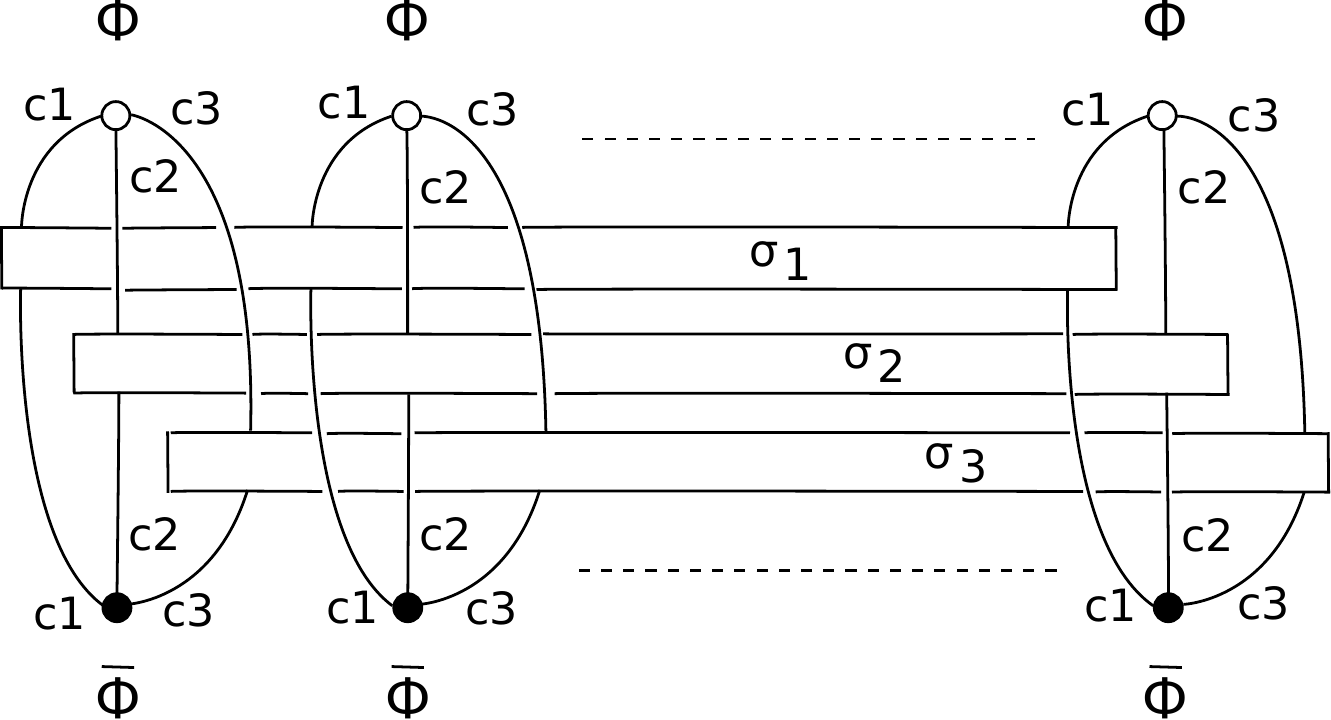}
\vspace{0.3cm}
\caption{ {\small  The contraction of $n$ tensors $\Phi_{c_1c_2c_3}$ and 
$n$ tensors $\bar\Phi_{c_1c_2c_3}$ identified as
permutation triple { $(\s_1,\s_2,\s_3)$}.}}
\label{countingC}
\end{minipage}
\end{center}
\end{figure}

All the  contractions  between the indices of   $n$ tensors  and $n$ conjugate tensors producing $U(N) \times U(N) \times U(N)$ invariants can be described by triples of permutations $\sigma_1, \sigma_2, \sigma_3 \in S_n$ depicted in Figure \ref{countingC}. 

Use Figure \ref{countingC}, and fixing a gauge $\sigma_3=\id$, we obtain tensor invariants from permutation pairs, where pairs related by conjugating with a diagonal  adjoint action on  $  {S}_n \times {S}_n$ are in the same equivalence class : 
 \be
( \tilde\s_1 , \tilde\s_2  ) \sim (\tau  \tilde\s_1 \tau^{-1} , \;  \tau  \tilde\s_2 \tau^{-1} \,) \,, \quad 
\quad  \tilde\s_i, \tau \in {S}_n\,.
\label{orbitad}
\ee
  These equivalence classes are also known to enumerate bi-partite ribbon graphs.  There is thus a graphical interpretation of the rank $3$ tensor invariants in terms of bipartite ribbon graphs.
 
Burnside's lemma allows us to write the number of equivalence classes under a group action 
in terms of the fixed points of the same action.  This leads us to 
\bea
\Rib(n)   =
  \sum_{ p \,\vdash n }  {\rm Sym} ( p ) \,, 
 \label{blem}
 \eea 
where the sum is performed over all partitions $p$ of $n$, denoted $p \,\vdash n$, 
and where  $ {\rm Sym} (p):= \prod_{i=1}^n (i^{p_i})(p_i!)$. 
There is another expression of the same counting as a sum 
over triples $ ( R_1 , R_2 , R_3) $ of irreducible representations (irreps)  \cite{JoSan2}: 
\bea 
\Rib(n)  = \sum_{ R_1 , R_2 , R_3 \vdash n } C ( R_1 , R_2 , R_3 )^2 
\eea
The Kronecker coefficient $ C ( R_1 , R_2 , R_3 )$ is a non-negative-integer 
that yields the number of times $R_3$ appears in the tensor product decomposition $ R_1 \otimes R_2$.

\ 

\noindent{\bf The algebra $\cK(n)$ of bipartite ribbon graphs.}
For the group action given in \eqref{orbitad}, we consider
the set of  orbits, each of which is  associated with a ribbon graph. 
We introduce a label $r \in \{ 1, \dots, \Rib(n)\}$.  

Consider $ \mC ( S_n ) \otimes_{ \mC }  \mC ( S_n )$,  simply denoted
$ \mC ( S_n ) \otimes \mC ( S_n ) $.
For each ribbon graph labeled by $r$, 
consider its orbit representative given by the pair 
$( \tau_1^{(r)} ,  \tau_2^{(r)} )$.  
The  element $E_r$ of $ \mC ( S_n ) \otimes \mC ( S_n ) $ 
is defined as 
\bea 
\label{classr}
E_r=
{ 1 \over n! }  \sum_{ \mu \in S_n } \mu \tau_1^{(r)}  \mu^{-1}  \otimes \mu \tau_2^{(r)}  \mu^{-1} 
\eea
Now, we define  the $\mC$-vector subspace $\cK(n) \subset  \mC ( S_n ) \otimes \mC ( S_n )$ spanned by these elements:   
\be
\cK(n) = {\rm Span}_{\mC}\Big\{    
E_r \,, r=1, \dots \Rib(n) 
\Big\}
\label{graphbasis}
\ee
The dimension of $\cK(n) $ is the number of bipartite ribbon graphs
\be \Dim ( \cK ( n ) )  = \Rib(n) 
\ee 
$ \cK(n)$ is a  permutation centralizer algebra (PCA) \cite{PCA2016}  - a subspace of a permutation algebra, here $ \mC ( S_n ) \otimes \mC ( S_n ) $, which commutes with  a sub-algebra with basis labelled by permutations, here the diagonally embedded $S_n$ permutations. 
$\cK(n) $ is also semi-simple: it has a  non-degenerate  symmetric bilinear pairing 
given  by 
\bea\label{Wpairing}  
\bdel_2 : \mC(S_n)^{\otimes 2}\times  \mC(S_n)^{\otimes 2} \to \mC 
\eea
which is defined in terms of  the usual delta function on the group.
 $\bdel_2 (  \otimes_{i=1}^2 \s_i ; \otimes_{i=1}^2 \s'_i ) = 
\prod_{i=1}^2 \delta (\s_i\s'_i)$,  with $\delta(\s)=1$, if $\s=\id$, and $0$ otherwise. 
This  extends to linear combinations with complex coefficients. 
Semi-simplicity implies that, by the Wedderburn-Artin theorem,   $\cK(n) $ admits a decomposition in simple matrix algebras.  This decomposition is made manifest using what we denote as the Fourier basis 
\bea\label{qbasis}
Q^{R_1,R_2,R_3}_{\tau_1,\tau_2} &=& 
\kappa_{R_1,R_2}
\sum_{\s_1, \s_2 \in S_n}
\sum_{i_1,i_2,i_3, j_1,j_2}
C^{R_1,R_2; R_3 , \tau_1  }_{ i_1 , i_2 ; i_3 } C^{R_1,R_2; R_3, \tau_2  }_{ j_1 , j_2 ; i_3 } 
\crcr
&\times& 
 D^{ R_1 }_{ i_1 j_1} ( \sigma_1  ) D^{R_2}_{ i_2 j_2 } ( \sigma_2 )  \,  \sigma_1 \otimes \sigma_2  
\eea
 $D^R_{ ij} ( \sigma ) $ are the matrix elements of 
the linear operator $D^R(\s)$ in an orthonormal basis for the irrep $R$. The indices 
$ \tau_1 , \tau_2 $ run over an orthonormal basis for the  multiplicity space of $R_3$ appearing in the tensor 
decomposition    of $ R_1 \otimes R_2$. This multiplicity is equal to the Kronecker coefficient $C ( R_1 , R_2 , R_3 )$.  $\kappa_{R_1,R_2} = \frac{d(R_1)d(R_2)}{(n!)^2}$ is a normalization factor, where 
$d(R_i)$ is the dimension of the irrep $R_i$.  $C^{R_1,R_2; R_3 , \tau_1  }_{ i_1 , i_2 ;i_3 } $ are Clebsch-Gordan coefficients of the representations of $S_n$. 

\

\noindent{\bf 
Quantum mechanics of  bipartite ribbon graphs.}
We define a sesquilinear form on $ \mC ( S_n ) \otimes \mC ( S_n )$ as 
\bea\label{def:innerprod}
g ( \sum_i a_i   \alpha_{1i}  \otimes  \alpha_{2i} ,    \sum_j  b_j  \beta_{1j} \otimes  \beta_{2j}) 
= \sum_{i,j} \bar  a_i   b_j  \; 
\delta (\alpha_{1i} ^{-1 }\beta_{1j}  ) \delta (   \alpha_{2i} ^{-1} \beta_{2j} ) 
\eea
where $a_i,b_i \in \mC$, $\alpha_{1i},  \alpha_{2i} ,\beta_{1j} ,  \beta_{2j} \in S_n$,  and  where the bar means complex conjugation. 
One checks that
$g$ is nondegenerate and therefore induces an inner product on $\mC ( S_n) \otimes \mC(S_n)$. 
We restrict $g$ to an give an inner  product on $\cK(n)\subset \mC ( S_n) \otimes \mC ( S_n) $, and consequently, $\cK(n)$ is an algebra which is also an Hilbert space. 

There is another operator that will be of prominent use in the following. 
Consider the linear conjugation operator $S: \mC(S_n) \to \mC(S_n)$ which  maps a linear combination 
$A = \sum_{i} c_i \s_i \in \mC(S_n)$ to  $S(A) := \sum_{i} c_i \s_i^{-1}$. 
Extend this operation to $\mC ( S_n )  \otimes \mC ( S_n) $ by inverting the permutation in each tensor factor, 
we  easily check that $S^2 = \id$ and call $S$ an involution.

Let us discuss the Hermitian operators in our setting
that could play the role of Hamiltonian operators. 
For a conjugacy class $ \cC_{\mu}  \subset  S_n $ labelled by $\mu \vdash n$, a partition of $n$,  we have a central element 
$
T_{ \mu } = \sum_{ \sigma \in \cC_{ \mu } } \sigma 
$
that obviously obeys $\g T_\mu \g^{-1}  = T_\mu$, for any $\g \in S_n$. 
We are interested  in particular $\mu= [k,1^{n-k}]$ defined by a single cycle of length $k$  and all remaining cycles of length 1. 
The corresponding operator denotes $T_k$.

There are operators
that multiplicatively generate the centre of $\cK(n)$ \cite{KR1911}. 
At any $n \ge 2$,  we define elements in   $\mC ( S_n) \otimes \mC(S_n)$  
\bea\label{tki}
T^{(1)}_k = T_k \otimes 1    \,,  \qquad 
T^{(2)}_k = 1 \otimes T_k    \,,  \qquad 
T^{(3)}_k = \sum_{ \sigma \in \cC_k }  \sigma \otimes  \sigma  \;. 
\eea
\vspace{-0.4cm}

The sum of products of the $T^{(i)}_k$'s, $k=1, \dots, n$, generates the centre
$\cZ (\cK(n))$ of $\cK(n)$. In fact, one does not need the entire set $k=1, \dots, n$
to generate the centre, only a smaller number of them is enough $k=1, \dots, k_*(n) \le n$. 
\vspace{-0.4cm}

We showed that  $T_k^{(i)} $ are Hermitian operators on $\cK(n)$
with respect to in the inner product defined by \eqref{def:innerprod}
$ g ( E_s , T_k^{ (i)} E_r )= g ( T_k^{ (i)} E_s , E_r )$. 
(See proof of Proposition 3 in \cite{BenGeloun:2020yau}.)
The operators $T_k^{(i)} $, for $i$ ranging over $\{ 1,2, 3 \}$ and $k$ ranging over some subset of $\{ 2,3, \cdots , n  \}$ form a set of commuting Hermitian operators on $\cK(n)$. The commutativity follows from the fact that they are central elements of $ \cK(n)$.  
We  consider such sets of operators as Hamiltonians and we define a quantum-mechanical time evolution of states in $\cK ( n ) $  of the form 
\bea 
E_r (t) = e^{ - i t T_{ k}^{(i)} } E_r 
\eea
where $E_r (t)$ becomes time-dependent ribbon graph states.

The action of the $T_{ k}^{(i)}$'s  on the ribbon graph bases yields a crucial 
fact. As linear operators, let $(\cM^{ (i)}_k )_r^s $ be the matrix elements of $T_k^{(i)}$. We have
\bea\label{Ter}
T_k^{(i)} E_s = \sum_{ s } (\cM^{ (i)}_k )_s^t  E_t 
\eea
The matrix elements $ (\cM^{ (i)}_k )_r^s $ are non-negative integers 
(Proposition  2 in  \cite{BenGeloun:2020yau}). 

 $T_k^{(i)} $ operators act on the Fourier basis of $ \cK(n)$
as (Proposition 4  \cite{BenGeloun:2020yau}) 
\vspace{-0.5cm}
 \begin{proposition}
   \label{propTaQo}
 For all $k \in \{ 2, 3, \cdots  n \}$,  $\{ R_i \vdash n : i \in \{  1,2,3\}  \} $, $\tau_1, \tau_2  \in  [\![1, C(R_1,R_2,R_3) ]\!]$,  the Fourier basis elements 
  $ Q^{R_1, R_2, R_3}_{\tau_1 , \tau_2}$ are   eigenvectors of $T_k^{(i)} $: 
\bea
 T_k^{(i)} Q^{R_1, R_2, R_3}_{ \tau_1 , \tau_2 } 
  = { \chi_{R_i} ( T_k ) \over d(R_i) }   Q^{R_1, R_2, R_3}_{ \tau_1 , \tau_2 } \,,
\label{t1Qo} 
\eea
 \end{proposition}
 \vspace{-0.4cm}
 This means that the Fourier basis  $ Q^{R_1, R_2, R_3}_{\tau_1 , \tau_2}$ 
 is an eigenbasis of the operators $T_k^{(i)}$. 
 Furthermore, following Proposition 5 of \cite{BenGeloun:2020yau}, 
 we have 
 \vspace{-0.5cm}
\begin{proposition} 
\label{PropTkFS} 
For any $\widetilde k_*  \in \{ k_*(n) , k_*(n) +1 , \cdots , n \} $ the list of eigenvalues of the 
reconnection operators 
$$ \{  T^{(1)}_{2} , T^{(1)}_{ 3} , \cdots , T^{(1)}_{ \widetilde k_* } ;  T^{(2)}_{2} , T^{(2)}_{ 3} , \cdots , T^{(2)}_{ \widetilde k_* }  ; T^{(3)}_{2} , T^{(3)}_{ 3} , \cdots , T^{(3)}_{ \widetilde k_*}   \}$$  
uniquely  determines the Young diagram triples $(R_1 , R_2 , R_3 )$.
\end{proposition} 
 \vspace{-0.4cm}
The  sum of all permutations $ \sigma $ in the conjugacy  class $ \cC_{ p } $ in $ S_n$ for partition 
$p $ are central elements in $ \cZ ( \mC ( S_n)) $. The irreducible normalized 
 characters of these central elements are integers : 
\bea \label{chiint}
 \chi_R ( T_{ p } ) / d ( R )    \in \mathbb{Z} 
\eea
The proof combines a known number theoretic fact about the normalized characters of  a finite group 
 being algebraic integers, along with the rationality of characters of irreducible representations of $ S_n$
 which follows from the Murnaghan-Nakayama Lemma. 
 
 \

\noindent{\bf Stacking $T_k^{(i)}$ matrices and common eigenspace.}
Using Proposition \ref{PropTkFS}, the Fourier subspace for a given triple $ ( R_1 , R_2 , R_3 )$ 
is uniquely specified as  common eigenspace of the operators $T^{(i)}_k$, for 
$k \in \{ 2, \dots, \tilde k_*(n) \} $ and $i\in \{  1,2,3 \} $; with $ k_*(n)  \le \widetilde k_* \le n  $, 
with specified eigenvalues for these reconnection operators.  
The numerator $ \chi_R   ( T_k)$ is given by $\chi_R   ( T_k) = {T_k } \chi_R ( \sigma ) $ for $ \sigma \in \cC_k$. The character $ \chi_R ( \sigma )$ can be computed with the combinatorial Murnaghan-Nakayama rule  \cite{MurnaghanOnReps}. The dimension $d(R)$ is obtained from the hook formula for dimensions. 

The vectors in the  Fourier subspace for a triple $(R_1, R_2 , R_3)$ solve the following matrix equation 

\bea\label{stack}
\scriptsize{ \left[\begin{array}{c}\
\cM^{(1)}_2  - { \chi_{R_1} ( T_2 ) \over d({R_1}) }  \\
\vdots \\
\cM^{(1)}_{\widetilde k_*} -  { \chi_{R_1} ( T_{\widetilde k} ) \over d({R_1}) } \\
(1 \rightarrow 2) \\ 
(1 \rightarrow 3) 
\end{array}\right] 
 \cdot v 
  = {\bold 0 } 
  }
\eea
where the notation $(1 \rightarrow j) $, $j=2,3$, 
means that we replace the matrix block $\cM^{(1)}_k \rightarrow \cM^{(j)}_k  $ for $j=2,3$. 
This rectangular array gives the matrix elements of a linear operator mapping  $ \cK(n)$ 
to $3 (\widetilde k_* -1 ) $ copies of  $ \cK(n) $,  using the geometric basis of ribbon graph vectors. From   \eqref{chiint}, the normalized characters are integers.  Renaming as 
$\mathcal{L}_{ R_1 , R_2 , R_3 } $ the integer matrix in  \eqref{stack} we have
\bea\label{Xv} 
\mathcal{L}_{ R_1 , R_2 , R_3 } \cdot v = 0 
\eea
 We then  have, for each triple of Young diagrams, the problem of finding the null space of an integer matrix. Null spaces of integer matrices have integer null vector bases. These  can be interpreted in terms of lattices and can be constructed via  well known algorithms.

\section{The Hermite Normal Forms algorithm and lattice interpretation of kernels}\label{HNF-algo} 

We are interested in solving the linear system $X \cdot v = 0$, 
where $X= \mathcal{L}_{ R_1 , R_2 , R_3 } $  \eqref{Xv}   has only integer matrix entries. 
A crucial fact about the null spaces  of integer matrices is that they have bases given as integer vectors. This follows from the theory of Hermite normal forms (HNF) and has an interpretation in terms of sub-lattices \cite{Cohen}\cite{Schrijver}. 
The null space of the integer matrix   $X$  is the span of a set of null vectors which can be chosen to be integer vectors, i.e. integral linear combinations of the $E_r$. A key result from the theory of integer matrices and lattices is that any integer matrix $A$ (square or rectangular; we use $ A = X^T$) has a unique HNF. This means that $ A $ has a  decomposition $A= U h$ with $U $ a unimodular matrix, i.e. an integer  matrix of determinant $ \pm 1$ and $h$ is an integer matrix computed by
an integrality perserving algorithm. 

 In  the present application we have a lattice $\mathbb{Z}^{   \Rib(n)   } \subset \IR^{  \Rib(n) } $
which is interpreted as the space of integer linear combinations of the geometric ribbon graph basis vectors $E_r$ of the ribbon graph algebra $ \cK(n)$. 
Based on these facts, we provide  the construction of $ C(R_1,R_2,R_3)^2 $ as the dimension of a sub-lattice of  the lattice of ribbon graphs. 

The HNF  procedure  achieves the proof of the  
following theorem (Theorem 1 in \cite{BenGeloun:2020yau}): 
 \vspace{-0.4cm}
\begin{theorem}\label{theo:C2}
For every triple of Young diagrams $(R_1 , R_2 , R_3 ) $ with $n$ boxes, the lattice  $\IZ^{ | \Rib(n) | }$ 
of integer linear combinations of the geometric basis vectors $E_r$ of $ \cK ( n ) $ contains a sub-lattice 
of dimension $ ( C ( R_1 , R_2 , R_3 ))^2 $ spanned by a basis of  integer null vectors of the operator 
$ X$, which is $ \mathcal{L}_{ R_1, R_2 , R_3 } $ from \eqref{Xv}. 
\end{theorem}
 \vspace{-0.4cm}
 
The action of operator $S$ (see section \ref{sect:CandAlg})  on the geometrical ribbon graph basis $E_r$ or on the Fourier basis $Q$ of $ \cK ( n )$ has key properties that will allow us to interpret the dimension of various lattice subspaces
of $\IZ^{ | \Rib(n) | }$ in terms of sums of products of Kronecker coefficients. 

Let us denote the  vector space of ribbon graphs, which is the underlying vector space of the algebra $ \cK ( n ) $ 
by $ V^{ \Rib(n) } $. $ V^{ \Rib(n) } $ has a decomposition according to the eigenvalues of $S$ 
\bea 
V^{ \Rib(n) } = V^{ \Rib(n) }_{ S=1} \oplus V^{ \Rib(n) }_{ S=-1} 
\eea 
The action of $S$ on the Fourier basis $Q$ leads to: 
\bea\label{VR1R2R3}  
V^{ \Rib(n) } = \bigoplus_{R_1 , R_2 , R_3} V^{ \Rib (n) :\; R_1 , R_2 , R_3 }  
\eea
where $ V^{ \Rib (n):\; R_1 , R_2 , R_3}   $ has dimension $ C ( R_1 , R_2 , R_3  )^2  $ 
 and is spanned by the $Q^{ R_1 , R_2 , R_3  }_{ \tau_1 , \tau_2 }$ for all $\tau_1$ and $\tau_2$. 
Then $ V^{ \Rib (n) :\; R_1 , R_2 , R_3 }=  V^{ \Rib (n) :\; R_1 , R_2 , R_3 } _{ S =1 }\oplus  V^{ \Rib (n) :\;R_1 , R_2 , R_3  } _{ S =-1 }$. 
Combining this with \eqref{VR1R2R3} we then have 
\be
V^{ \Rib  ( n ) } = \bigoplus_{ R_1 , R_2 , R_3 }  \left (  V^{ \Rib ( n ) : R_1 , R_2 , R_3 }_{ S =1 } \oplus V^{ \Rib ( n ) : R_1 , R_2 , R_3 }_{ S =-1 }  \right ) 
\ee
We can show that 
\bea 
\Dim  \left ( V^{ \Rib ( n ) : R_1 , R_2 , R_3 }_{ S =-1 }  \right )  
&=&  { C ( R_1 , R_2 , R_3) ( C ( R_1 , R_2 , R_3 ) -1 ) \over 2 }  \crcr
& = & \Dim  \left ( P^{ R_1 , R_2 , R_3 } V^{ \Rib (n ) }_{ \pairs^-} \right ) 
\label{dim-}
\eea
  with $P^{ R_1 , R_2 , R_3 }$ the projector onto $V^{ \Rib ( n ) : R_1 , R_2 , R_3 }$ and $V^{ \Rib(n) }_{ S=-1} =  V^{ \Rib ( n ) }_{ \pairs^- } $.   Similarly,   we can show   
\bea 
&&
 \Dim \left ( V^{ \Rib ( n ) : R_1 , R_2 , R_3 }_{ S =+1 } \right )   =  { C ( R_1 , R_2 , R_3 ) ( C ( R_1 , R_2 , R_3 )  + 1 ) \over 2 } \cr 
& & =    \Dim  \left ( P^{ R_1 , R_2 , R_3 } V^{ \Rib (n ) }_{ \pairs^+}  \right )  + \Dim 
\left (  P^{ R_1 , R_2 , R_3 } V^{ \Rib (n ) }_{ \singlets} \right )   
\label{dim+sing}
\eea
  with the decomposition 
  $V^{ \Rib(n) }_{ S=1} = V^{ \Rib ( n ) }_{ \pairs^+ }  \oplus V_{ \singlets } $, that  defines $ V^{ \Rib ( n ) }_{ \pairs^+ }  $, the subspace spanned by $\{ ( E_r^{(n)}  + E_r^{ (\bar n )}  ) \}$, 
   and $ V_{ \singlets }  $ the subspace spanned by $\{E_r^{ (s)} \}$ that are ribbons obeying $SE_r^{ (s)} = E_r^{ (s)}$. 
Note that we do not have separate expressions for the two terms in the sum above 
in terms of Kronecker coefficients, since we do not expect the $P^{ R_1 , R_2 , R_3 }$ to  commute with the projection of $V^{ \Rib(n) }_{ S=1}$ into  the separate summands 
 $V^{ \Rib(n) }_{ \singlets }$ and $V^{ \Rib(n) }_{ \pairs^+ }$.

 Using once again the HNF procedure 
 and  the outcome of the above discussion, 
 we reach the statement (Theorem 2 in \cite{BenGeloun:2020yau})
  \vspace{-0.3cm}
 \begin{theorem}
 \label{addLatt}
For every triple of Young diagrams $(R_1 , R_2 , R_3 ) $ with $n$ boxes, there are three
constructible sub-lattices of $\IZ^{ \Rib(n) }$  of respective dimensions 
${ C ( R_1 , R_2 , R_3 ) ( C ( R_1 , R_2 , R_3 )  +1) /2 } $, 
${ C ( R_1 , R_2 , R_3 ) ( C ( R_1 , R_2 , R_3 )  -1) / 2 } $, 
and $C ( R_1 , R_2 , R_3 )$. 
 \end{theorem}
  \vspace{-0.4cm}
 
If we perform the sum over $ R_1 , R_2 , R_3 $  in \eqref{dim+sing}, we have 
$
 \Dim \left ( V^{ \Rib ( n ) }_{ S =+1 } \right ) 
= \Dim  \left ( V^{ \Rib ( n ) }_{ \pairs^+ }  \right )  + \Dim \left ( V_{ \singlets } \right )
$ 
and 
$
\Dim \left ( V^{ \Rib ( n ) }_{ S = -1  } \right )
= \Dim  \left ( V^{ \Rib ( n ) }_{ \pairs^- }  \right )$. 
Since $
\Dim  \left ( V^{ \Rib ( n ) }_{ \pairs^+ }  \right ) = \Dim  \left ( V^{ \Rib ( n ) }_{ \pairs^- }  \right )
$,  we have 
 \bea\label{sumsinglets}  
&& \hbox{ Number of bi-partite  ribbon graphs with $n$ edges invariant under   } S \cr 
&&  =  \Dim  \left ( V^{ \Rib ( n ) }_{ \singlets }  \right ) = \sum_{ R_1 , R_2 , R_3 }   C ( R_1 , R_2 , R_3 )  
\eea 
While the sum over triples of Young diagrams with $n$ boxes of the square of Kronecker coefficients gives 
the number of ribbon graphs with $n$ edges, the sum of the Kronecker coefficients gives the number of singlet ribbon graphs. 
 
\ 

\begin{center} 
{ \bf Acknowledgments} 
\end{center} 
The research of S.R. was supported by the STFC consolidated grant ST/P000754/1 “ String Theory, Gauge Theory and Duality”. 
 We thank the Editors Konstantinos Anagnostopoulos, Peter Schupp, George Zoupanos, for the invitation to contribute to this special volume on ``Non-commutativity and physics’’.  
 We thank Peter Schupp for suggestions on the introduction.

\vskip.2cm 
 
%
%

\end{document}